\documentclass[11pt]{article}
\pdfoutput=1

\usepackage[pdftex,linktoc=all]{hyperref}

\usepackage[utf8]{inputenc}
\usepackage[T1]{fontenc}
\usepackage{amsmath}
\usepackage{cite}
\usepackage{graphicx,wasysym}
\usepackage{slashed}
\usepackage{soul,xcolor}
\usepackage{color}
\usepackage{simplewick}

\newcommand{\p}{\partial}
\newcommand{\lb}{\bar{\lambda}}
\newcommand{\la}{\lambda}

\newcommand{\s}{\sigma}

\newcommand{\be}{\begin{equation}}
\newcommand{\ee}{\end{equation}}
\newcommand{\bea}{\begin{align}}
\newcommand{\eea}{\end{align}}

\def\det{\mathop{\rm det}}

\def\Label#1{\label{#1}
\smash{\hbox to0pt{\raise1ex\hbox{\tiny[#1]}\hss}}}

\numberwithin{equation}{section}
\begin{document}
\frenchspacing
$\qquad$\vspace{2.5cm}
\begin{center}
{ \Large {\bf
Tree amplitudes from nonlinear (super)symmetries}}
\vspace{0.3cm}
\\
{ \Large {\bf in Volkov--Akulov theory}}
\vspace{1cm}

Anna Karlsson$^{\flat}$, Hui Luo$^{\natural}$ and Divyanshu Murli$^{\flat}$

\vspace{1cm}

{\small
$^{\flat}${\it Stanford Institute for Theoretical Physics and Department of Physics, \\ Stanford University, Stanford, CA 94305, U.S.A.}}

\vspace{0.5cm}

{\small
$^{\sharp}${\it II. Institut f\"ur Theoretische Physik, Universit\"at Hamburg, \\ Luruper Chausee 149, 22761 Hamburg, Germany }}

\vspace{1cm}

\end{center}
\abstract{The characteristics of tree level scattering amplitudes in theories with nonlinear (super)\-symmetries were recently proposed by Kallosh to be encoded in a simple way directly from the action, based on a background field method. We check this conjecture to lowest non-trivial order in the Volkov--Akulov theory, where the local $6$-point diagram cancels and the results up to $6$ points indeed are in agreement with S-matrix computations.}
\vfill
\noindent\makebox[\linewidth]{\rule{\textwidth}{0.4pt}}
{\footnotesize \texttt{e-mails: annakarl@stanford.edu, hui.luo@desy.de, divyansh@stanford.edu}}
\thispagestyle{empty}

\newpage
\pagenumbering{arabic}
\tableofcontents
\vspace{0.1cm}
\noindent\rule{\linewidth}{0.4pt}

\section{Introduction}
Nonlinear symmetries are of interest e.g. in the setting where constrained superfields give rise to Volkov--Akulov (VA) models \cite{Volkov:1973ix}, since those symmetries are useful for describing features within cosmology, such as inflation \cite{Antoniadis:2014oya,Ferrara:2014kva,Ferrara:2015tyn,Carrasco:2015pla}. For example, de Sitter supergravity is characterised by a local version of VA supersymmetry \cite{Bergshoeff:2015tra,Hasegawa:2015bza,Kuzenko:2015yxa,Bandos:2015xnf}. The properties inferred by the symmetries are central to understanding these models, and one question concerns what characterises the scattering amplitudes.

Amplitudes in theories with nonlinear (super)symmetries can be determined e.g. through Feynman rules (introduced for VA in \cite{Chen:2014xoa})\footnote{Amplitude expressions in coordinate space can be found in \cite{Liu:2010sk}.}, recursion relations as in \cite{Luo:2015tat}, and the Cachazo--He--Yuan (CHY) representation \cite{Cachazo:2013hca,He:2016vfi,Cachazo:2016njl}. In each setting, the symmetries of the S-matrix are represented in different ways, in part giving rise to the question of what the general effects of nonlinear symmetries of the action on the S-matrix are. Recently, this problem was investigated in \cite{Kallosh:2016qvo}, giving rise to a conjecture for a generating functional of the S-matrix in terms of the tree level diagrams. 

Generating functionals for scattering amplitudes have been studied before, e.g. for gauge and gravity theories in \cite{Rosly:1996vr,Rosly:1997ap,Selivanov:1997aq,Selivanov:1998hn} a method which applies more widely, and relations to the background field method are mentioned in \cite{Srednicki:2007qs}. However, the relation between the method of generating functionals and nonlinear symmetries remains to be understood. Here, the approach in \cite{Kallosh:2016qvo} represents an extension to theories with global nonlinear symmetries in general, rather than for specific examples. This makes the conjecture in \cite{Kallosh:2016qvo} interesting to investigate further, not only for its potential use in understanding the symmetries of the amplitudes in question, but also to see if it may highlight the connection between generating functionals and nonlinear symmetries.

The conjecture of \cite{Kallosh:2016qvo} is that the $n$-point tree amplitudes $A_{i_1\ldots i_n}$ should be encoded in
\be\label{eq.SmS'}
\left.\left(S(\phi)-S_{,i}\phi^i\right)\right|_{\phi=\phi[\phi_0]}=\sum_{n=4}^\infty\frac{1}{n!} A_{i_1\ldots i_n}\phi_0^{i_1}\ldots\phi_0^{i_n}\,,
\ee
with $S_{,i}=\delta S/\delta \phi^i$ and a solution to the set of a fields $(\phi^i)$ given by a method inspired by the background field method in the formalism of DeWitt \cite{DeWitt:1967ub,Kallosh:1974yh,Grisaru:1975ei}. Note that $\phi$ is the background field, with the classical field equation
\be
\frac{\delta S}{\delta\phi^i}=\phi^j_0 \vec{S}_{,ji}^0\,,\quad S,_{ij}^0 G_0^{jk}=-\delta_k^i\,,
\ee
with $G_0^{ij}$ a free propagator. Moreover, $\phi^i$ includes all fields of the given model and (in DeWitt's formalism) there is a summation convention such that repeated indices include integrations over the spacetime coordinates, implied in \eqref{eq.SmS'}. $\phi_0$ is a general solution of the free field equation and $\phi[\phi_0]$ encodes the interactions through
\be\label{eq.bckgSol}
\phi^i[\phi_0]\equiv\phi_0^i+G_0^{ij}\sum_{n=2}^\infty\frac{1}{n!}t_{j i_1\ldots i_n}\phi_0^{i_1}\ldots\phi_0^{i_n}\,,
\ee
obtained \emph{by iteration} from
\be\label{eq.prop}
\phi^i=\phi^i_0 +G_0^{ij}\frac{\delta S^\text{int}(\phi)}{\delta\phi^j}\,.
\ee
In a theory of interactions, $\phi[\phi_0]\neq \phi_0$, which is crucial in \eqref{eq.SmS'}.

The relation \eqref{eq.SmS'} is especially interesting in that it represents a prediction of the amplitudes based on the symmetries of the action, rather than observations of the symmetries of the amplitudes once they have been constructed. In construction, it represents a generalisation to nonlinear symmetries originating in global symmetry breaking of a method known to describe properties of tree-level S-matrix elements in non-Abelian gauge theories and gravity. As such, its foundation is sound, yet the crucial question of whether or not the conjecture \eqref{eq.SmS'} indeed holds, remains. A priori, this is not guaranteed. As mentioned previously, the connection between generating functionals and nonlinear symmetries (in general) is an open question. 

The objective of this paper is to analyse the consistency of the conjecture \eqref{eq.SmS'}, in the simplest non-trivial setting. For this, we look to the Volkov--Akulov (VA) model \cite{Volkov:1973ix}, which encodes nonlinear symmetries. The VA model is comparatively simple, relevant in the setting of investigating amplitudes, and represents a type of nonlinear symmetry not previously investigated in relation to generating functionals. Moreover, its characteristics are well-known. In specific, the local $6$-point vertex is known to vanish. It vanishes in S-matrix computations, and there is a field redefinition (nonlinear change of variables) which maps the VA theory onto the Komargodski--Seiberg (KS) action \cite{Casalbuoni:1988xh,Komargodski:2009rz}, where the absence of the $6$-point vertex is manifest. We will not discuss the different merits of the two dual theories, useful as they are in different settings\footnote{The KS action has no terms with six spinors, but does have terms with eight spinors \cite{Kuzenko:2010ef}, while the opposite  is true for the VA action \cite{Kuzenko:2005wh}.}, but one thing is clear --- in the VA theory, the absence of the $6$-point vertex is a product of the symmetries in the theory in a non-trivial way, and this must be captured by the relation in \eqref{eq.SmS'}. Hence, the simplest non-trivial check (relevant in the amplitude setting discussed earlier) of \eqref{eq.SmS'} is the analysis of what the left hand side (LHS) gives for the VA $6$-point amplitude. For the generating functional to work at the lowest non-trivial order, the LHS must encode the vanishing of the local vertex at that order, through the symmetries of the VA theory.

The paper is organised as follows. In section 2 the relation \eqref{eq.SmS'} is analysed up to $6$-point interactions in free fields, beginning with the LHS and ending with the subsequent predictions for the $4$- and $6$-point amplitudes, where the local $6$-point term indeed vanishes, indicating that the correct physics is captured. We then proceed to make comparisons with the known results for the aforementioned tree level amplitudes, in section 3, noting an agreement between the methods, also confirmed by direct S-matrix computations. Finally, we discuss how the cancellation of the VA local $6$-point vertex comes about through the generating functional in \eqref{eq.SmS'}.

\section{Pure Volkov--Akulov theory up to $6$-point interactions}
For a check of \eqref{eq.SmS'} to lowest order in the VA model ($\mathcal{N}=1,\,D=4$), we are interested in the VA action up to $6$-point interactions. Here, $(\phi^i)$ corresponds to the VA goldstino $(\psi^a,\bar\psi^{\dot a})$, and we use the normalisation in \cite{Kuzenko:2005wh,Kuzenko:2010ef},
\be\label{eq.ourVA}
S_\text{VA}[\psi,\bar\psi]=-\frac{1}{2\kappa^2}\int\mathrm{d}^4x \det\left({\delta_\mu}^\nu+i\kappa^2\psi\stackrel{\leftrightarrow}{\partial}_\mu\sigma^\nu\bar\psi\right)\,.
\ee
To lowest order in fields, up to the local $6$-point term and with $\kappa^2=1$, this is
\begin{subequations}
\begin{align}
S_2&=i\int\mathrm{d}^4x\, (\partial_\mu\psi)\sigma^\mu\bar\psi\,,\label{eq.S2}\\
S_4&=-2\int\mathrm{d}^4x\, (\psi\sigma^{[\mu}\p_\mu\bar\psi)(\p_\nu\psi\sigma^{\nu]}\bar\psi)\,,\label{eq.S_4}\\
\left.S_6\right|_{(\psi,\bar\psi)=(\la,\lb)}&=-\frac{i}{2}\int\mathrm{d}^4x\,(\la\s^b\p_a\lb)(\la\s^c\stackrel{\leftrightarrow}{\p_b}\lb)(\p_c\la\s_a\lb)\,,\label{eq.S_6}
\end{align}
\end{subequations}
where $(\la,\lb)$ represent the free fields $(\phi^i_0)$ subject to the equations
\be\label{eq.freeEOM}
-i(\la\stackrel{\leftarrow}{\slashed{\p}})_{\dot a}=0\,,\quad-i(\slashed{\p}\lb)_a=0\,.
\ee
The solution to \eqref{eq.bckgSol} given in \cite{Kallosh:2016lwj} for $\psi$ is easily extended to $\bar\psi$, also to lowest order in perturbation,
\begin{subequations}\label{eqs.fieldexps}
\begin{align}
\psi&=\la+i\frac{\delta S^\text{int}(\psi,\bar\psi)}{\delta \bar\psi}\stackrel{\leftarrow}{\slashed{\p}^{-1}}\,,\qquad \left.\frac{\delta S_4}{\delta \bar\psi}\right|_{(\psi,\bar\psi)=(\la,\lb)}=-(\la\s^\mu\stackrel{\leftrightarrow}{\p}_\nu\lb)(\p_\mu\la\s^\nu)\,,\\
\bar{\psi}&=\lb+i{\slashed{\p}^{-1}}\frac{\delta S^\text{int}(\psi,\bar\psi)}{\delta \psi}\,,\qquad \left.\frac{\delta S_4}{\delta \psi}\right|_{(\psi,\bar\psi)=(\la,\lb)}=-(\s^\nu\p_\mu\lb)(\la\s^\mu\stackrel{\leftrightarrow}{\p}_\nu\lb)\,,
\end{align}
\end{subequations}
with the notation
\be\label{eq.25}
(\la\sigma\stackrel{\leftrightarrow}{\p}\lb)=(\la\sigma\p\lb)-(\p\la\sigma\lb)\,.
\ee
Note that $(\psi,\bar\psi)$ in \eqref{eqs.fieldexps} needs to be solved iteratively to get the correct answer to all orders, i.e. by replacing $\psi\rightarrow(\la+\ldots)$ in $\delta S^\text{int}/\delta\bar\psi$, etc. Also, $\frac{\slashed{\p}}{\p^2}\slashed{\p}\la\neq0$ --- the free equation of motion cannot be employed when propagators are present.

For spinor conventions, we use
\be \label{eq.symConv}
\s^{(\mu\nu)}=\frac{1}{2}\left(\s^\mu\s^\nu+\s^\nu\s^\mu\right)=-\eta^{\mu\nu}\,
\ee
and the Fierz identities
\be\label{eq.Fierz}
(\la\sigma^\mu\bar\eta)(\la\sigma^\nu\bar\eta')=\frac{1}{2}(\eta\sigma^\mu\sigma^\nu\eta')\la^2\,,\quad (\la\sigma^\mu\bar\eta)(\la'\sigma^\nu\bar\eta)=\frac{1}{2}(\la\sigma^\mu\sigma^\nu\la')\bar\eta^2\,.
\ee
We have $\la\sigma\lb=-\lb\sigma\la$, keeping $\bar\sigma$ implicit.

\subsection{Amplitudes from the action}
To begin with, we note that
\be\label{corr.1}
S_{,i}\phi^i=\psi^i\frac{\partial S}{\partial \psi^i}+\bar\psi^i\frac{\partial S}{\partial \bar\psi^i}
\ee
is required for \eqref{eq.SmS'} to give the correct result in the VA theory. We analyse \eqref{eq.SmS'} up to the $6$-point contributions in free fields, with
\begin{gather}\begin{aligned}\label{eq.rightSide}
\big(S[\psi]-S_{,i}\psi^i\big)=&-\frac{1}{4}\int\mathrm{d}^4x\, (\la\s^\nu\stackrel{\leftrightarrow}{\p}_\mu\lb)(\la\s^\mu\stackrel{\leftrightarrow}{\p}_\nu\lb)\\
&-i\int\mathrm{d}^4x\, (\la\s^\mu\stackrel{\leftrightarrow}{\p}_\nu\lb)(\p_\mu\la\s^\nu \stackrel{\rightarrow}{\left(\frac{\p_\delta}{\p^2}\right)}\s^\delta\s^\rho\p_\tau\lb)(\la\s^\tau\stackrel{\leftrightarrow}{\p}_\rho\lb)\\
&+i\int\mathrm{d}^4x\, \la^2(\p_\mu\lb\p_\nu\lb)(\p^\nu\la\s^\mu\lb)\\
&+\mathcal{O}(\la^4\lb^4)\,,
\end{aligned}\end{gather}
as described below. Note that the $4$-point vertex equals $S_4(\la,\lb)$, and that the third term is $S_6(\la,\lb)$.

\subsubsection*{Contribution from $S_2$}
The contribution to \eqref{eq.SmS'} from \eqref{eq.S2} is
\be\label{eq.S211}
\big(S_2[\psi]-(S_2)_{,i}\psi^i\big)=i\int\mathrm{d}^4x\, \psi\slashed{\p}\bar\psi
\ee
up to total derivatives, and contributes to both the $4$- and $6$-point amplitudes. In part, the contributions follow a simple pattern. In inserting \eqref{eqs.fieldexps}, with one $(\la,\lb)$ in the expansion the contribution is
\be\label{eq.S21st}
-\left.\left(\la\slashed{\p}\slashed{\p}^{-1}\frac{\delta S^\text{int}}{\delta \psi}+\frac{\delta S^\text{int}}{\delta \bar\psi}\stackrel{\leftarrow}{\slashed{\p}^{-1}}\slashed{\p}\lb\right)\right|_{(\psi,\bar\psi)=(\la,\lb)}=\left[\phi^i=(\psi,\bar\psi)\right]=\phi^i_0 \frac{\delta S^\text{int}(\phi^i)}{\delta \phi^i}
\ee
There are two parts to this expression. One with $(\psi,\bar\psi)=(\la,\lb)$,
\be
\left.(4S_4+6S_6)\right|_{(\psi,\bar\psi)=(\la,\lb)}\,,
\ee
and a second where an additional correction from $S_4$ in \eqref{eqs.fieldexps} is of relevance, since amplitudes up to $6$ points are under consideration. This extra correction is distributed on the three fields $(\la,\lb)$ of \eqref{eqs.fieldexps}, but due to the form of \eqref{eq.S21st}, this effectively corresponds to an extra insertion on each field in $3S_4$, the same as considering the $6$-point contribution of
\be\label{eq.3S4}
\left.3S_4(\psi,\bar\psi)\right|_\text{$6$-pt}\,.
\ee

In addition, we have a final contribution from inserting \eqref{eqs.fieldexps} into \eqref{eq.S211}
\be
-i\frac{\delta S_4}{\delta \bar\psi}\slashed{\p}^{-1}\left.\frac{\delta S_4}{\delta \psi}\right|_{(\psi,\bar\psi)=(\la,\lb)}=-i\int\mathrm{d}^4x\, (\la\s^\mu\stackrel{\leftrightarrow}{\p}_\nu\lb)(\p_\mu\la\s^\nu \stackrel{\rightarrow}{\left(\frac{\p_\delta}{\p^2}\right)}\s^\delta\s^\rho\p_\tau\lb)(\la\s^\tau\stackrel{\leftrightarrow}{\p}_\rho\lb)\,,\label{eq.S2D6}
\ee
which corresponds to a symmetric $6$-point diagram with a propagator between two $3$-point vertices. The derivative in the middle denotes an overall derivative, acting on all of the fields to the right of it.

\subsubsection*{Contributions from $S_4$ and $S_6$}
Equation \eqref{eq.S_4}
gains a factor of $-3$ through \eqref{eq.SmS'}. This gives a $4$-point interaction
\begin{subequations}
\be
\left.-3S_4\right|_{(\psi,\bar\psi)=(\la,\lb)}=-3(\la\s^\mu\p_\nu\lb)(\p_\mu\la\s^\nu\lb)=\frac{3}{4}(\la\s^\mu\stackrel{\leftrightarrow}{\p}_\nu\lb)(\la\s^\nu\stackrel{\leftrightarrow}{\p}_\mu\lb)\,,
\ee
and a $6$-point interaction
\be
-3(\la\s^\mu\stackrel{\leftrightarrow}{\p}_\nu\lb)\left[\p_\mu\la\s^\nu\left.\bar\psi\right|_{\mathcal{O}(\la\lb^2)}-\left.\psi\right|_{\mathcal{O}(\la^2\lb)}\s^\nu\p_\mu\lb\right]\,,
\ee
\end{subequations}
which is $-6$ times the contribution from $S_2$, shown in \eqref{eq.S2D6}. However, this contribution directly cancels the one from \eqref{eq.3S4}.

The local $6$-point term from $S_6$ acquires a factor of $-5$ through \eqref{eq.SmS'}. Moreover, the free equation of motion, the Fierz identity and $\la^3=0$ can be used to simplify \eqref{eq.S_6} into
\be\label{eq.loc6}
\left.S_6\right|_{(\psi,\bar\psi)=(\la,\lb)}=i\int\mathrm{d}^4x\, \la^2(\p_\mu\lb\p_\nu\lb)(\p^\nu\la\s^\mu\lb)\,.
\ee
In total, this gives \eqref{eq.rightSide}.

\subsection{Tree amplitude diagrams}
Reconnecting to the LHS of \eqref{eq.SmS'}, the $4$-point part of \eqref{eq.rightSide} is
\be\label{eq.lad2la}
-\int\mathrm{d}^4x\,(\la\s^\nu\p_\mu\lb)(\la\s^\mu\p_\nu\lb)\stackrel{\eqref{eq.Fierz}}{=}
\frac{1}{2} \int\mathrm{d}^4x\,\la^2 \p^2\lb^2\,.
\ee
Recall that $\p^2\la=0$ since the expressions are modulus the free equation of motion.

In momentum space, with $\partial=i\hat p$ and the amplitude conventions of \cite{Chen:2014xoa}
\be
\langle ij\rangle\equiv (\la'_i\la'_j)\,,\quad[ij]\equiv (\tilde\la'_i\tilde\la'_j)\,,\quad s_{ij}\equiv(p_i+p_j)^2\,,
\ee
with even numbers for $\la$ and odd for $\lb$, the \eqref{eq.SmS'} prediction for $A_4$ is
\be\label{eq.A4}
A_4=2s_{13}\langle24\rangle[13]\,.
\ee
Here, \eqref{eq.lad2la} has been multiplied by $-4$. The sign is to ensure the enumeration of the fields to be compatible with ordering from S-matrix contractions, as after (10) in \cite{Chen:2014xoa}, where a positive enumeration corresponds to the choice in \eqref{eq.choice}. This is required for the results to be comparable, and should be done after all rearrangements of the spinors, as the amplitude and vertex definitions are different. The translation between the two descriptions includes $(\la_a,\lb_{\dot a})\rightarrow(\la'_a,\tilde\la'_{\dot a})$ and sign corrections for permutations from a positive ordering from the right,
\be\label{eq.forder}
(\la_4\la_2)(\lb_3\lb_1)\rightarrow-\langle42\rangle[31]=\langle42\rangle[13]\,.
\ee
The factor of $4$ is to capture the permutations $1\leftrightarrow3$ and $2\leftrightarrow4$, symmetries already present through
\be
\la^2_{ij}=\la^2_{(ij)}\,,\quad \lb^2_{ij}=\lb^2_{(ij)}\,.
\ee
In \eqref{eq.SmS'} this correcting factor shows as $n!$ ($4!$ rather than $4$) but there the LHS includes permutations over all of the free fields, disregarding the separation between the two chiralities. For consistency, we take the right hand side (RHS) of \eqref{eq.SmS'} to encode an effective removal of the permutations over the different channels, with $n!$ replaced by $(n/2)!(n/2)!$ in the VA model,
\be\label{corr.2}
\eqref{eq.SmS'}\,:\quad \sum_{n=4}^\infty\frac{1}{n!} A_{i_1\ldots i_n}\phi_0^{i_1}\ldots\phi_0^{i_n}
\quad\stackrel{\text{VA}}{\longrightarrow}\quad\sum_{k=2}^\infty\frac{1}{k!k!} A_{i_1\ldots i_{k}j_1\ldots j_{k}}\la^{i_1}\ldots\la^{i_k}\lb^{j_1}\ldots\lb^{j_k}\,.
\ee

\subsubsection*{The $6$-point diagram, with cancellation}
The second term in \eqref{eq.rightSide} is
\be\label{eq.6pt}
i\int\mathrm{d}^4x\,\left[(\p^2\la^2)\lb-\frac{1}{2}(\la^2)\lb\stackrel{\leftarrow}{\p^2}\right]\frac{\slashed{\p}}{\p^2}\left[\la(\p^2\lb^2)-\frac{1}{2}\stackrel{\rightarrow}{\p^2}\la(\lb^2)\right]\,.
\ee
Here we have used
\begin{gather}\begin{aligned}
(\la\s^\mu\stackrel{\leftrightarrow}{\p}_\nu\lb)(\p_\mu\la\s^\nu)_\alpha=+&(\p^2\la^2)\lb_\alpha-\frac{1}{2}\left[(\la^2)\lb_\alpha\right]\stackrel{\leftarrow}{\p^2}\\
&+\left[(\la\s^\mu\lb)(\p_\nu\la\s^\nu)_\alpha\right]\stackrel{\leftarrow}{\p_\mu}
\end{aligned}\end{gather}
and the corresponding for $(\s^\rho\p_\tau\lb)_\beta(\la\s^\tau\stackrel{\leftrightarrow}{\p}_\rho\lb)$, which differs by a sign, while noting that the third term only is nonzero when coupled to $\slashed{\p}^{-1}$, corresponding to a matching with the first term, and never gives a contribution to \eqref{eq.rightSide} in that combination.

In \eqref{eq.6pt}, the cross terms vanish due to the free equation of motion \eqref{eq.freeEOM}:
\begin{gather}\begin{aligned}\label{eq.crossterms}
&(\p^2\la^2)\left[(\lb\s^\mu\p_\mu\la)\lb^2+(\lb\s^\mu\la)\p_\mu\lb^2\right]\\
=\,&2(\p^2\la^2)(\lb\s^\mu\la)(\lb\p_\mu\lb)\propto(\p^2\la^2)\lb^2(\la\s^\mu \p_\mu\lb)=0\,.
\end{aligned}\end{gather}

The two last terms give
\be\label{eq.2.26}
-i\la^2(\p_\mu\lb\p_\nu\lb)(\p^\nu\la\s^\mu\lb)\,,
\ee
which cancels against the contribution from $S_6$, the last term in \eqref{eq.rightSide}. This represents the important cancellation of the local $6$-point vertex.

The remaining term is
\begin{gather}\begin{aligned}\label{eq.2.27}
-i(\p^2\lb^2)\la\stackrel{\leftarrow}{\left(\frac{\slashed{\p}}{\p^2}\right)}\lb(\p^2\la^2)\,.
\end{aligned}\end{gather}
Using the same procedure as for $A_4$ for the \eqref{eq.SmS'} prediction for $A_6$, this should be multiplied by the number of diagrams summing over $(135)$ and $(246)$: 36 (rather than $6!$) and corrected by a sign due to ordering, which in combination gives the nine terms
\be\label{eq.A6}
A_6=4\langle24\rangle[35]\langle6|3+5|1]\frac{s_{24}s_{35}}{s_{124}}+ (\text{cyclic in 1,3,5 and 2,4,6})\,.
\ee

\section{Comparisons with other amplitude approaches}
In comparing results in the VA theory, possible to describe using different actions and different normalisations, mappings between the settings have to be taken into consideration. In the previous section, we used \eqref{eq.ourVA}. The KS action \cite{Casalbuoni:1988xh,Komargodski:2009rz}
\be\label{eq.KS}
S_{KS} =\int\mathrm{d}^4x\, \left(-iG\slashed{\p}\bar G+\frac{\kappa^2}{2}\bar G^2\p^2G^2-\frac{\kappa^6}{2}G^2\bar G^2\p^2G^2\p^2\bar G^2\right)
\ee
corresponds to \eqref{eq.ourVA} through a nonlinear relation $\psi(G,\bar G)$ \cite{Kuzenko:2010ef,Luo:2009ib}. Here, the absence of the local 6-point vertex is manifest, which makes it especially convenient for analyses up to $6$-point interactions.

With a different normalisation of the VA action, $\kappa^2\neq1$ in \eqref{eq.ourVA}, the parts of the action change as
\be\label{eq.VA'}
S'_{2n}=(\kappa^2)^{n-1}\left(S_{2n})\right|_{\kappa^2=1}\,.
\ee
This also alters the field expansion in free fields in \eqref{eqs.fieldexps}, in the same way at each order in fields, giving the general expression
\be\label{eq.Aconv}
A'_{2n}=(\kappa^2)^{n-1}\left(A_{2n})\right|_{\kappa^2=1}\,.
\ee
This can also be seen directly from that each $4$-point vertex is accompanied by a difference amounting to a factor of $\kappa^2$ (and $\kappa^4$ for local $6$-point vertices), giving \eqref{eq.Aconv} for tree diagrams.

\subsubsection*{The $4$- and $6$-point amplitudes}
In comparing with amplitude results obtained using recursion relations in \cite{Luo:2015tat}, the CHY representation in \cite{Cachazo:2016njl} and one instance of S-matrix computations in \cite{Chen:2014xoa}, we note that some of the results are given with the normalisation choice $\kappa^2=1/2$. As a consequence, the amplitude predictions from the previous section, in \eqref{eq.A4} and \eqref{eq.A6}, need to be reinterpreted using \eqref{eq.Aconv}.

The $4$-point amplitude of \cite{Chen:2014xoa} is precisely the $A_4$ in \eqref{eq.A4}. The amplitude in \cite{Luo:2015tat,Cachazo:2016njl}, with $\kappa^2=1/2$, is
\be
A'_4=s_{13}\langle24\rangle[13]\,,
\ee
in agreement with the expression from \eqref{eq.SmS'}. The same is true for the $6$-point amplitude \cite{Luo:2015tat,Cachazo:2016njl}
\be\label{eq.A'6}
\qquad A'_6=\langle24\rangle[35]\langle6|3+5|1]\frac{s_{24}s_{35}}{s_{124}}+(\text{cyclic in 1,3,5 and 2,4,6})\,.
\ee

\subsection{Feynman tree amplitudes}
A double-check of the comparison between the amplitude derivations can be made through deriving the $4$- and $6$-point amplitudes directly from the KS action in \eqref{eq.KS}. If done for the VA action \eqref{eq.ourVA} the S-matrix procedure simply gives an expression equivalent to the LHS of \eqref{eq.SmS'}, i.e. \eqref{eq.rightSide}, before contraction with external particles. In this way, \eqref{eq.SmS'} reduces to the same expression as the S-matrix result. For a second check, the interaction Hamiltonian for the KS action is
\begin{eqnarray}\label{Hamil-KS}
H_I= \int \mathrm{d}^4 x \, \left(-\kappa^2\overline G^2 \partial_\mu G\partial^\mu G-\kappa^2\overline G^2\, G\partial^2 G+{\kappa^6\over 2}G^2 \overline G^2 \partial^2 G^2 \partial ^2\overline G^2\right)\,, 
\end{eqnarray}
and the non-trivial part of the S-matrix ($\mathcal S= 1+i\, \mathcal T$) can be computed from
\be
\langle p_1\dots p_n |i\,\mathcal T| p'_1 \dots p'_k \rangle = \lim_{T\rightarrow \infty (1-i \epsilon)} \,_0\langle p_1\dots p_n | T \exp \left[-i \int_{-T}^T \mathrm{d}t\, H_I(t)\right] | p'_1 \dots p'_k \rangle_0\,.
\ee

\subsubsection*{The 4-point amplitude}
To derive the $4$-point amplitude (with $\kappa^2=1$), it is sufficient to consider only incoming particles
\be\label{eq.choice}
\,_0 \langle 0|i\,\mathcal T| p_1, p_2, p_3, p_4\rangle_0 =\,_0\langle 0 |T\left( i \int \mathrm{d}^4 x \, \overline G^2 \partial_\mu G\partial^\mu G \right)| p_1, p_2, p_3, p_4\rangle_0\,,
\ee
where all the particles are treated as incoming particles. The term containing $\partial^2 G$ in the four-point interaction vanishes due to the external fields being on-shell ($p^2=0$), and Wick's theorem gives that only normal ordered operators contribute. Following the conventions in appendix \ref{app.F}, we derive the four-point amplitude of two initial left-handed fermions $(\psi_2,\psi_4)$ and two initial right-handed fermions $(\overline\psi_1,\overline\psi_3)$. For this scattering process, \eqref{eq.choice} gives four terms
\begin{gather}\begin{aligned}
&\,_0\langle 0|T \bigg( i \int \mathrm{d}^4 x \, \bcontraction{}{\overline G} {\,\,\overline G \partial_\mu G \,\partial^\mu G \bigg)|p_1, p_2,}{p_3} \overline G
\bcontraction[2ex]{}{\overline G}{\partial_\mu G \,\partial^\mu G \bigg)| }{p_1} \,\, \overline G \partial_\mu \contraction {}{ G}{\,\partial^\mu G \bigg)| p_1, p_2, p_3,}{p_4}
 \contraction[2ex]{G\,\partial^\mu} {G}{\bigg)| p_1,} {p_2} G\,\partial^\mu G \bigg)| p_1, p_2, p_3, p_4\rangle_0+( p_1\leftrightarrow p_3,\, p_2\leftrightarrow p_4 )\\
&= (2\pi)^4 \,\delta^{(4)}(p_1+p_2 + p_3 +p_4) \cdot i\, \mathcal M(\overline \psi_1,\psi_2,\overline \psi_3,\psi_4)\,,
\end{aligned}\end{gather}
where all possible equivalent contractions are counted in, i.e. with $G$ to $p_2$ or $p_4$ and $\overline G$ to $p_1$ and $p_3$. The amplitude with these specific initial and final states is
\begin{eqnarray}
A_4=\mathcal M(\overline \psi_1,\psi_2,\overline \psi_3,\psi_4) =2 s_{13} \langle24\rangle [13]. \label{4pt-FR}
\end{eqnarray}

\subsubsection*{The 6-point amplitude}
Since there is no $6$-point interaction in the Hamiltonian \eqref{Hamil-KS}, the $6$-point amplitude only receives contributions from the $4$-point interactions
\begin{gather}\begin{aligned}\label{eq.6step1}
\,_0 \langle 0 |T\bigg[{i^2\over 2! } &\int \mathrm{d}^4 x \, (\overline G^2 \partial_\mu G\partial^\mu G+\overline G^2\, G\partial^2 G) \\
\,\,\times&\int \mathrm{d}^4 y \, (\overline G^2 \partial_\mu G\partial^\mu G+\overline G^2\, G\partial^2 G) \bigg]| p_1 \ldots p_6 \rangle_0,
\end{aligned}\end{gather}
where, again, all particles are treated as incoming ($\sum_{i=1}^6 p_i =0$). This can be manipulated as described in appendix \ref{app.F}, in total giving the full 6-point amplitude
\begin{align}\label{eq.F6pt}
A_6&=\mathcal M(\overline \psi_1,\psi_2,\overline \psi_3,\psi_4, \overline \psi_5, \psi_6)\nonumber\\
&= 4 \langle24\rangle [35]\langle 6|3+5|1] { s_{24}s_{35}\over s_{124}}+(\text{cyclic in 1,3,5 and 2,4,6})\,,
\end{align}
in agreement with \eqref{eq.A6}.

\section{Discussion}
We have checked the internal consistency of the method for determining tree level scattering amplitudes in theories with nonlinear (super)symmetry suggested in \cite{Kallosh:2016qvo}, amounting to \eqref{eq.SmS'}, to lowest order in the VA model. The conjectured relation, representing a specific generalisation of the background field method to nonlinear global symmetries, reproduces the $4$- and $6$-point amplitudes well-known from e.g. \cite{Chen:2014xoa,Luo:2015tat,Cachazo:2016njl}, with the modifications \eqref{corr.1}, \eqref{corr.2} and signs due to ordering of the fields as in \eqref{eq.forder}. At higher orders, signs may appear in the comparisons, since the conventions used for the metric and the propagator differ from what is typically used in amplitude calculations. Importantly, the vanishing of the local $6$-point vertex represents a non-trivial result since the check is done in the traditional VA action, and not the dual KS action where its absence is manifest. As such, the conjecture in \eqref{eq.SmS'} is quite likely to capture the correct physics of the amplitude diagrams.

The separate way of determining the amplitudes through the method inspired by the background field method works in that the background field solutions substituted into the action provide the same structure as S-matrix derivations. Basically, the solutions of the fields in terms of free fields give rise to an overcounting of the terms contributing to the amplitudes, when introduced in the action alone. This is amended by the additional consideration of $S_{,i}$, which alters the relative factors between the parts of the action $(S_n)$. This simplification happens in a comparatively roundabout manner --- \eqref{eq.SmS'} is not a convenient way to determine amplitudes compared with S-matrix computations. The symmetry \eqref{eq.SmS'} represents is however interesting in what it says about the structure of the amplitudes, and it would be interesting to see what would characterise a corresponding description for loops. Beyond tree level, further non-local effects are expected, in a general expression corresponding to \eqref{eq.SmS'}.

As to the relation between the generating functional in \eqref{eq.SmS'} and the nonlinear VA symmetries, implied by the RHS of \eqref{eq.SmS'}, those symmetries are implicit in the connections between the terms with different numbers of spinors in the action, i.e. the $S_n$. The key to the nonlinear symmetries is how the structures of $S_n$ and $S_{n+2}$ are related. In the amplitude setting, this is exemplified in cancellations between diagrams originating from different $S_n$. While local $6$-point terms are absent in an analysis of the KS action, the case of the VA theory presents a cancellation between the $S_6$ term and connected $4$-point vertices. Herein lies a difference, and the generating functional in \eqref{eq.SmS'} captures the relative symmetries, relevant for the cancellation to take place. Considering the VA generating functional in \eqref{eq.rightSide}, the key feature is the relative coefficients of the 6-point terms (the second and third term on the RHS), as well as the reduction of the second term into the two parts \eqref{eq.2.26} (local) and \eqref{eq.2.27} (non-local). In fact, the third term is just $S_6$ with free fields, so the interesting part from a symmetry point of view is how the generating functional gives the second term in \eqref{eq.rightSide}, with the crucial local part. It does this by using the structure of the action itself. In a sense, \eqref{eq.SmS'} is partly a shell correcting relative coefficients, and partly an iterative procedure giving terms built on the interactions through the solution for the background field in \eqref{eq.prop}. The latter is what produces the interesting $6$-point term, as shown in \eqref{eq.S2D6}, where two $S_4$ vertices (both products of $\phi[\phi_0]$) are connected by a propagator. Hence, while the generating functional in \eqref{eq.SmS'} does not contain the specific symmetries, it does encode iterative interactions of the theory, which in the end has the desired effects --- the consequences of the nonlinear symmetries.

\section*{Acknowledgements}
We thank R.~Kallosh for helpful discussions. This work is supported by the SITP and the US National Science Foundation Grant PHY-1316699. AK is also supported by the Knut and Alice Wallenberg Foundation, and HL by the German Science Foundation (DFG) within the Collaborative Research Center 676 `Particles, Strings and the Early Universe'.

\appendix
\section{Feynman tree amplitudes --- conventions and 6-point derivation}\label{app.F}
In the S-matrix derivations, we use $\eta^{\mu \nu}={\rm diag}(-1,+1,+1,+1)$ and $\sigma^\mu_{\alpha \dot \alpha}$ and $(\overline \sigma^\mu)^{\dot \alpha \alpha}$:
\begin{gather}\begin{aligned}
&\sigma^0 = \overline \sigma^0= \left(\begin{array}{cc} 1 ~~& 0\\ 0 ~~& 1 \end{array}\right)\,, ~~\quad \sigma^1 =- \overline \sigma^1= \left(\begin{array}{cc} 0 ~~& 1\\ 1 ~~& 0 \end{array}\right)\,,\\
&\sigma^2 =- \overline \sigma^2= \left(\begin{array}{cc} 0 ~~& -i\\ i ~~& 0 \end{array}\right)\,, \quad \sigma^3 =- \overline \sigma^3= \left(\begin{array}{cc} 1 ~~& 0\\ 0 ~~& -1 \end{array}\right)\,.
\end{aligned}\end{gather}
Quantized fermionic fields are written as 
\begin{gather}\begin{aligned}\label{field-Fourier}
&G_\alpha (x) = \int {\mathrm{d}^3 p\over (2\pi)^3} {1\over \sqrt {2E_p}} \sum_s \left(u_\alpha^- (p,s) a(p,s) e^{i p\cdot x} + v_\alpha^- (p, s) a^\dagger(p,s) e^{-i p\cdot x}\right)\,,\\
&\overline G_{\dot \alpha} (x) = \int {\mathrm{d}^3 p\over (2\pi)^3} {1\over \sqrt {2E_p}} \sum_s \left(v_{\dot \alpha}^+(p,s) a(p,s) e^{i p\cdot x} + u_{\dot \alpha}^+(p, s) a^\dagger (p,s) e^{-i p\cdot x}\right)\,,
\end{aligned}\end{gather}
where the creation and annihilation operators obey the anti-commutation
\begin{eqnarray}
\{a(p,s) ,\,a^\dagger(p^\prime,s^\prime)\} = \delta^3 (p-p^\prime) \delta_{s\,s^\prime}\,.
\end{eqnarray}
For massless particles, $u_\alpha^- (p,s)$, $ v_\alpha^- (p, s)$, $u_{\dot \alpha}^+(p,s)$ and $v_{\dot \alpha}^+(p,s)$ satisfy
\be
(p\cdot \overline \sigma)^{\dot \alpha \alpha} u^-_\alpha = (p\cdot \overline \sigma)^{\dot \alpha \alpha} v^-_\alpha = 0\,, \quad u^+_{\dot \alpha} (p\cdot \overline \sigma)^{\dot \alpha \alpha} = v^+_{ \dot \alpha} (p\cdot \overline \sigma)^{\dot \alpha \alpha} = 0\,,
\ee
indicating $v_\alpha^- (p, s)=u_\alpha^- (p,s)\equiv\la'_\alpha(p,s)$ and $v_{\dot \alpha}^+(p,s) = u_{\dot \alpha}^+(p,s)\equiv\tilde\la'_{\dot\alpha}(p,s)$. The summations over spins are $\sum_s u^-_\alpha(p,s)\, u^+_{\dot\alpha}(p,s) = p\cdot \sigma_{\alpha \dot\alpha} $ and $\sum_s u^{+\dot \alpha}(p,s)\, u^{- \alpha}(p,s) = p\cdot \overline \sigma^{\dot\alpha \alpha}$.

Covariant normalization of the one-particle states $|p,s\rangle \equiv (2\pi)^3 \sqrt{2 E_p} \,a^\dagger(p,s) |0\rangle$ is used, and the
two-component external state spinors are assigned to be the following
\begin{itemize}\setlength\itemsep{0em}
\item[---] Initial states (incoming) are left-handed fermions: $G_\alpha(x)| p, s\rangle\sim \la'_\alpha(p,s)$, 
\item[---] Initial states (incoming) are right-handed fermions: $\overline G_{\dot \alpha}(x)| p, s\rangle\sim \tilde\la'_{\dot\alpha}(p,s)$,
\item[---] Final states (outgoing) are left-handed fermions: $\langle p, s|\overline G_{\dot \alpha}(x)\sim \tilde\la'_{\dot \alpha}(p,s)$, 
\item[---] Final states (outgoing) are right-handed fermions: $\langle p, s|G_\alpha(x)\sim \la'_{\alpha}(p,s)$.
\end{itemize}
With the definition in \eqref{eq.prop}, the propagators are
\begin{gather}\begin{aligned}
&\contraction {}{G}{_\alpha(x)\,\,} {\overline G} G_\alpha(x) \,\,\overline G_{\dot\alpha} (y)= -\int {\mathrm{d}^4p\over (2\pi)^4} { i p\cdot \sigma_{\alpha \dot\alpha} \over p^2+ i\epsilon} e^{ip\cdot (x-y)}\,, \\
&\contraction {}{\overline G}{^{\dot\alpha}(x)\,\,} { G} \overline G^{\dot\alpha}(x) \,\,G^{\alpha}(y) = -\int {\mathrm{d}^4p\over (2\pi)^4} { i p\cdot \overline \sigma^{\dot\alpha \alpha} \over p^2+ i\epsilon} e^{ip\cdot (x-y)}\,.
\end{aligned}\end{gather}
This differs from the most common conventions, where $\eta^{\mu \nu}={\rm diag}(+1,-1,-1,-1)$ (incoming particles with $e^{-ipx}$ and vice versa) and $\slashed{\p}\slashed{\p}^{-1}=\delta$ rather than $-\delta$, the latter which gives rise to an overall sign in the equation above.

\subsection*{The $6$-point derivation}
Denoting $ V_{41} = \overline G^2 \partial_\mu G\partial^\mu G$ and $V_{42} = \overline G^2\, G\partial^2 G$, \eqref{eq.6step1} consists of the sum of four terms
\begin{eqnarray}\label{6-pt-original}
~_0 \langle 0 |T\left({i^2\over 2! } \int \mathrm{d}^4 x \, V_{4i}(x) \int \mathrm{d}^4 y \,V_{4j}(y)\right)|p_1, p_2, p_3, p_4, p_5, p_6 \rangle_0
\end{eqnarray}
with $(i,j)$ equal to $(1,1)$, $(1,2)$, $(2,1)$ and $(2,2)$. The term with two $V_{42}$ vanishes due to $p^2=0$ for external particles. The term with two $V_{41}$ can be computed using Wick's theorem to obtain
\begin{align}
&\,_0 \langle 0 |T\left({i^2\over 2! } \int \mathrm{d}^4 x \, \overline G^2 \partial_\mu G\partial^\mu G \,\,\int d^4 y \,\overline G^2 \partial_\mu G\partial^\mu G \right)| p_1\ldots p_6 \rangle_0\\
=\,\,&2 i^2 \int \mathrm{d}^4 x\,\mathrm{d}^4 y \left(\,_0 \langle 0 | : \overline G^2(x) \partial_\mu G^\alpha(x)\partial^\mu \contraction {}{G}{_\alpha(x)}{\overline G} G_\alpha(x) \overline G_{\dot\alpha}(y) \overline G^{\dot \alpha}(y) \partial_\nu G(y)\partial^\nu G(y) : | p_1\ldots p_6\rangle_0\right)\nonumber\\
+&2 i^2 \int \mathrm{d}^4 x\,\mathrm{d}^4 y \left(\,_0 \langle 0 | : \partial_\mu G(x)\partial^\mu G(x) \overline G_{\dot\alpha}(x) \contraction {}{\overline G}{^{\dot\alpha}(x)\partial_\nu}{G} \overline G^{\dot\alpha}(x) \partial_\nu G^{\alpha}(y) \partial^\nu G_\alpha(y)\overline G^2(y) : | p_1\ldots p_6 \rangle_0\right)\nonumber
\end{align}
where $:~:$ denotes normal ordering. Both lines in the above give the same result, and the combined factor of four in each line comes from four possible internal Wick contractions. The different contractions of $G$ or $\overline G$ with the incoming states give different helicity configurations, and following the 4-point amplitude derivation we analyse the scattering process with half of the initial states left- and right-handed, respectively. That is, $G$ contracts with the incoming particles with momenta $p_2,p_4,p_6$ to form left-handed states while $\overline G$ contracts with the incoming $p_1,p_3,p_5$ to form right-handed states. As a result, the contribution from the first term of \eqref{6-pt-original} consists of 9 terms
\begin{gather}\begin{aligned}\label{eq.int1}
\mathcal M(\overline \psi_1,\psi_2,\overline \psi_3,\psi_4, \overline \psi_5, \psi_6) \bigg|_{\rm T_1}
=& 16 [35] \langle24\rangle \langle 6|p_3+p_5|1] {(p_2\cdot p_4) (p_6\cdot (p_1+p_2+p_4))\over (p_1+p_2+p_4)^2}\\
&+(\text{cyclic in 1,3,5 and 2,4,6})\,.
\end{aligned}\end{gather}

The contribution from $(i,j)=(1,2)$ in \eqref{6-pt-original} is
\begin{eqnarray}
&&\,_0 \langle 0 |T\left({1\over 2! } i^2 \int \mathrm{d}^4 x \, \overline G^2 \partial_\mu G\partial^\mu G \,\,\int \mathrm{d}^4 y \,\overline G^2\, G\partial^2 G\right)| p_1\ldots p_6\rangle_0\nonumber\\
&=&i^2 \int \mathrm{d}^4 x\, \mathrm{d}^4 y \,_0 \langle 0|: \partial_\mu G(x) \partial^\mu G(x) \overline G_{\dot \alpha}(x) \contraction{}{\overline G}{^{\dot \alpha}(x) \,\, \partial^2_y} {G}\overline G^{\dot \alpha}(x) \,\, \partial^2_y G^\alpha(y) G_\alpha(y)\, \overline G^2(y):| p_1\ldots p_6 \rangle_0\,\,,\nonumber
\end{eqnarray}
which results in
\begin{gather}\begin{aligned}
\mathcal M(\overline \psi_1,\psi_2,\overline \psi_3,\psi_4, \overline \psi_5, \psi_6) \bigg|_{\rm T_2} 
= &4 \langle24\rangle {(p_2\cdot p_4)}\,\, [35] \langle 6|p_2+p_4|1] \\
&+ (\text{cyclic in 1,3,5 and 2,4,6})\,.
\end{aligned}\end{gather}
Eventually, this expression is zero, which is quite easy to see --- it corresponds to three sets of the cross terms in \eqref{eq.crossterms}. The contribution from $(i,j)=(2,1)$ is identical, so neither part contributes to the 6-point amplitude. Moreover, with this, \eqref{eq.int1} can be rearranged into an expression equal to \eqref{eq.F6pt}, giving the full 6-point amplitude.

\providecommand{\href}[2]{#2}\begingroup\raggedright\endgroup

\end{document}